\documentclass[showpacs,amssymb,amsmath,aps,prl,reprint,superscriptaddress]{revtex4-1}
\usepackage{graphicx}
\usepackage{bm}
\usepackage{dsfont}

\newcommand\I{\texttt{i}}

\newcommand\sech{\mathop{\rm sech}\nolimits}
\renewcommand\Re{\mathop{\rm Re}\nolimits}
\renewcommand\Im{\mathop{\rm Im}\nolimits}

\begin{document}
\title{Bright and Dark Solitons on the Surface of Finite-Depth Fluid Below the Modulation Instability Threshold}
\author{I.\,S.~Gandzha}
\author{Y\lowercase{u}.\,V.~Sedletsky}
\affiliation{Institute of Physics, Nat.~Acad.~of~Sci.~of Ukraine, Prosp.~Nauky 46, Kyiv 03028, Ukraine}
\email{gandzha@iop.kiev.ua, sedlets@iop.kiev.ua}

\date{\today}

\begin{abstract}
We use the high-order nonlinear Schr\"{o}dinger equation (NLSE)
derived to model the evolution of slowly modulated wave trains with
narrow spectrum on the surface of ideal finite-depth fluid. This
equation is the finite-depth counterpart of celebrated Dysthe's
equation, which is usually used for the same purpose in the case of
infinite depth. We demonstrate that this generalized equation admits
bright soliton solutions for depths below the modulation instability
threshold $kh\approx 1.363$ ($k$ being the carrier wave number and
$h$ the undisturbed fluid depth), which is not possible in the case
of standard NLSE. These bright solitons can exist along with the
dark solitons that have recently been observed in a water wave tank
[Phys. Rev. Lett. \textbf{110}, 124101 (2013)].
\end{abstract}
\maketitle

Bright and dark solitons are the fundamental self-localized modes of
the optical field in nonlinear dispersive media such as waveguides,
optical fibers, and photonic crystals \cite{Kivshar_Agrawal_2003}.
Bright solitons are characterized by a localized intensity peak on a
homogeneous background, while dark solitons can be described by a
localized intensity hole on a continuous wave background
\cite{Kivshar_1998}. Solitons are known to exist due to the balance
of dispersion and nonlinearity and propagate without changing their
shape and keeping their energy \cite{Dodd_solitons}. Bright solitons
are formed when the group-velocity dispersion in an optical fiber is
anomalous (or, similarly, when the nonlinearity of a planar
waveguide is self-focusing). In this case, the uniform carrier wave
is unstable with respect to long-wave modulations allowing for the
formation of solitons. This type of instability is known as the
modulation instability \cite{Zakharov_MI_2009}. On the contrary,
dark solitons are formed in the case of normal group-velocity
dispersion in fibers (or a self-defocusing nonlinearity in
waveguides), when a uniform carrier wave is modulationally stable.

In the case of water waves, bright solitons are known to appear in
the form of surface envelopes of modulated wave trains when the
uniform carrier wave is modulationally unstable \cite{Debnath_1994}.
This happens for water depths $h$ above the modulation instability
threshold, namely, at $kh > 1.363$, $k$ being the carrier wave
number. In addition to theoretical predictions, envelope solitons
were observed experimentally in
Refs.~\cite{Yuen_Lake_1975,Ablowitz_2000,Ablowitz_2001,Slunyaev_PF_2013},
mostly in the case of deep water ($kh\gg 1$). Dark solitons can
appear on shallow water below the modulation instability threshold,
at $kh < 1.363$ \cite{Peregrine_1983}. They have recently been
observed in a series of experiments performed in a water wave tank
\cite{Chabchoub_PRL_2013_Dark,Chabchoub_PRE_2014_Gray}.

In mathematical terms, bright and dark solitons are described by the
nonlinear Schr\"{o}dinger equation (NLSE) of the focusing and
defocusing types, respectively \cite{Debnath_1994,Kivshar_1998,Grimshaw_2011}.
NLSE takes into account the second-order dispersion and the phase
self-modulation (cubic nonlinear term). In the general context of
weakly nonlinear dispersive waves, this equation was first discussed
by Benney and Newell \cite{Benney_Newell_1967}. In the case of
gravity waves propagating on the surface of infinite-depth
irrotational, inviscid, and incompressible fluid, NLSE was first
derived by Zakharov \cite{Zakharov_1968}. The finite-depth NLSE was
first derived by Hasimoto and Ono~\cite{Hasimoto_Ono_1972}.

To achieve a better comparison with experiment and to trace
nonlinear effects in the vicinity of the modulation instability
threshold $kh \approx 1.363$, where the leading nonlinear term
vanishes, high-order nonlinear and nonlinear-dispersive effects
should be taken into consideration. In the case of infinite depth,
such a high-order NLSE (HONLSE) was first derived by Dysthe
\cite{Dysthe_1979} and then rewritten in a more general form by
Trulsen {\em et al.} \cite{Trulsen_Dysthe_2000}. It includes the
high-order dispersion and cubic nonlinear dispersive terms as well
as an additional nonlinear dispersive term describing the input of
the wave-induced mean flow. Dysthe's equation was extensively used
in numerical simulations of wave evolution and showed good agreement
with experiment and simulations based on fully nonlinear Euler
equations
\cite{Lo_Mei_1985,Akylas_1989,Clamond_2006,Slunyaev_PRE_2013,Chabchoub_2013_PRL_MNLS,Shemer_2013}.

In the case of finite depth, HONLSE is generally coupled with
additional equation for the wave-induced mean flow
\cite{Brinch-Nielsen_1986,Gramstad_Trulsen_JFM_2011}. No modeling of
water wave evolution has been performed with these general equations
because of their complexity. Sedletsky \cite{SedletskyJETP2003} used
an additional power expansion of the induced mean flow to derive a
single HONLSE for the first-harmonic envelope of surface profile.
This equation is the direct counterpart of Dysthe's equation
\cite{Dysthe_1979,Trulsen_Dysthe_2000} but for the case of finite
depth. Slunyaev \cite{Slunyaev_2005} confirmed the results obtained
in Ref.~\cite{SedletskyJETP2003} and extended them to the next
order. Gandzha {\em et al.} \cite{UJP_2014} rewrote this equation in
dimensionless form and used it to model the evolution of bright
solitons on finite depths. When the HONLSE terms were taken into
consideration, one-soliton solutions to NLSE were transformed into
quasi-soliton solutions with slowly varying amplitude. These
quasi-solitons were found to propagate with nearly constant speed
and possess the unique property of solitons to exist over long
periods of time without breaking. Their speed was found to be higher
than the speed of the bright NLSE solitons taken as initial
conditions in computations. This phenomenon was observed earlier
both in experiment and numerical modeling in the case of deep-water
limit in Refs.~\cite{Su_1982,Akylas_1989}.

In the present work we demonstrate that the HONLSE derived in
Ref.~\cite{UJP_2014} admits exact solutions in the form of bright
solitons below the modulation instability threshold. Such bright
solitons can be observed in the same experimental setup that was
used by Chabchoub {\em et al.}~\cite{Chabchoub_PRL_2013_Dark} to
observe dark solitons at $kh = 1.2$.

\begin{figure}[t]
\includegraphics[width=\columnwidth]{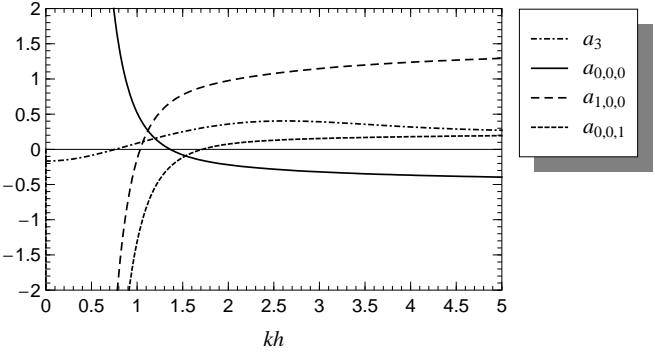}
\caption{\label{fig:an} HONLSE coefficients as functions of $kh$.}
\end{figure}

We start from the fully nonlinear Euler equations written for the
potential two-dimensional waves on the surface of irrotational,
inviscid, and incompressible fluid under the influence of gravity.
Waves are assumed to propagate along the horizontal $x$-axis
($-\infty< x<\infty$), and the direction of the vertical $y$-axis is
selected opposite to the gravity force. The fluid is assumed to be
bounded by a solid flat bed $y=-h$ at the bottom and a free surface
$y = \eta(x,\,t)$ at the top. The atmospheric pressure is assumed to
be constant on the free surface. Then the evolution of waves and
associated fluid flows is governed by the following set of equations
\cite{Stoker_1992}:
\begin{eqnarray}
\label{eq:Lapl}
 \Phi_{xx}+\Phi_{yy}=0,\;\; && -h\leqslant y \leqslant\eta(x,\,t);
\\ \label{eq:Dyn}
 \Phi_{t}+\frac{1}{2}\bigl(\Phi_{x}^2+\Phi_y^2\bigr)+g\eta=0,\;\; && y=\eta(x,\,t);
\\ \label{eq:Kin}
 \eta_{t}-\Phi_y+\eta_{x}\Phi_{x}=0,\;\; && y=\eta(x,\,t);
\\ \label{eq:Bottom}
 \Phi_y=0,\;\; && y=-h.
\end{eqnarray}
Here, $\Phi(x,\,y,\,t)$ is the velocity potential, $g$ is
acceleration due to gravity, $t$ is time, and the indices designate
the partial derivatives over the corresponding variables. The linear
dispersion relation for wave trains with carrier frequency $\omega$
and wave number $k$ is
\begin{equation}\label{eq:dispersion}
\omega^2 = gk\tanh(kh)\equiv gk\sigma,\;\;\sigma\equiv\tanh(kh),
\end{equation}
and the carrier group speed is
\begin{equation}\label{eq:Vg}
V_g=\frac{\partial \omega}{\partial k}=\frac{\omega}{2k\sigma}\Bigl((1-\sigma^2)kh+\sigma\Bigr).
\end{equation}
Considering slowly modulated wave trains with narrow spectrum around
the carrier frequency and wave number, the unknown free-surface
displacement and velocity potential can be looked for in the form of
Fourier series with slowly variable coefficients:
\begin{gather}
\biggl(\!\!\begin{array}{c}\eta(x,\,t)\\\Phi(x,\,y,\,t)\end{array}\!\!\biggr)=
\sum_{n=-\infty}^{\infty}\biggl(\!\!\begin{array}{c}\eta_n(x,\,t)\\\Phi_n(x,\,y,\,t)\end{array}\!\!\biggr)\exp\bigl(\I n(kx-\omega t)\bigr),\nonumber\\
\eta_{-n}\equiv \eta_n^*,\;\;\Phi_{-n}\equiv \Phi_n^*,\label{eq:Fourier}
\end{gather}
where $^*$ stands for complex conjugate. In the small-amplitude
approximation the unknown functions $\eta_n$ and $\Phi_n$ can be all
expressed in terms of the first-harmonic amplitude $\eta_1$ with the
use of additional multi-scale power expansions
\cite{SedletskyJETP2003}. In this way, the original set of equations
(\ref{eq:Lapl})--(\ref{eq:Bottom}) can be reduced to one evolution
equation for the complex-valued amplitude $\eta_1(x,\,t)$.
Introducing dimensionless time, coordinate, and amplitude
\begin{equation}
\tau = \frac{\omega}{c}\, t,\quad \chi = kx,\quad u = 2\sqrt{c}\,k\,\eta_1,
\end{equation}
this evolution equation can be written in the following form
\cite{UJP_2014}:
\begin{multline}\label{eq:HONLSE}
u_\tau = - a_1 u_\chi - \I a_2 u_{\chi\chi}+\I a_{0,\,0,\,0}\,|u|^2 u \\
+ \Bigl(a_3 u_{\chi\chi\chi} - a_{1,\,0,\,0}\,u_{\chi}|u|^{2} - a_{0,\,0,\,1}\,u^{2}u^{*}_{\chi}\Bigr).
\end{multline}
This HONLSE describes the evolution of the first-harmonic envelope
of the surface profile with taking into account the third-order
dispersion and cubic nonlinear dispersive terms. Here, $c$ and $a_1$
are the dimensionless phase and group speeds, respectively,
\begin{gather}
c = -\frac{4\sigma^2}{\upsilon}>0,\\
a_1 =\frac{ck}{\omega}V_g = -\frac{2\sigma}{\upsilon}\Bigl(\left(1-\sigma^2\right)kh+\sigma\Bigr)>0,\\
\upsilon = \left(\sigma^2-1\right)\left(3\sigma^2+1\right)k^2h^2-2\sigma\left(\sigma^2-1\right)kh-\sigma^2.\nonumber
\end{gather}
The second- and third-order dispersion coefficients are
\begin{gather}
a_2=\frac{1}{2},\quad a_3 = \frac{1}{12\sigma\upsilon}\sum_{p=0}^3 a_3^{(p)}(kh)^p,\\
a_3^{(0)}=-3\sigma^3,\quad a_3^{(1)}=-3\sigma^2\left(\sigma^2-1\right),\nonumber\\
a_3^{(2)}=-3\sigma\left(\sigma^2-1\right)\left(3\sigma^2+1\right),\nonumber\\
a_3^{(3)}=\left(\sigma^2-1\right)\left(15\sigma^4-2\sigma^2+3\right),\nonumber
\end{gather}
and the coefficient at the cubic nonlinear term is
\begin{gather}
a_{0,\,0,\,0} = -\frac{1}{16\sigma^4\nu}\sum_{p=0}^2 a_{0,\,0,\,0}^{(p)}(kh)^p,\\
a_{0,\,0,\,0}^{(0)} = -\sigma^2\left(7\sigma^4-38\sigma^2-9\right),\nonumber\\
a_{0,\,0,\,0}^{(1)} = 2\sigma\left(3\sigma^6-23\sigma^4+13\sigma^2-9\right),\nonumber\\
a_{0,\,0,\,0}^{(2)} = \left(\sigma^2-1\right)^2\left(9\sigma^4-10\sigma^2+9\right),\nonumber\\
\nu =\left(\sigma^2-1\right)^2 k^2 h^2 - 2\sigma\left(\sigma^2+1\right)kh + \sigma^2.\nonumber
\end{gather}
The cubic nonlinear dispersion coefficients are
\begin{gather}
\biggl(\!\!\begin{array}{c}a_{1,\,0,\,0}\\a_{0,\,0,\,1}\end{array}\!\!\biggr)=
\frac{1}{32\sigma^5\nu^2}\sum_{p=0}^5 \Biggl(\!\begin{array}{c}a_{1,\,0,\,0}^{(p)}\\a_{0,\,0,\,1}^{(p)}\end{array}\!\Biggr)(kh)^p,\\
a_{1,\,0,\,0}^{(0)} = \sigma^5\left(\sigma^6-40\sigma^4+193\sigma^2+54\right),\nonumber\\
a_{0,\,0,\,1}^{(0)} = -\sigma^5\left(\sigma^6-7\sigma^4+7\sigma^2-9\right),\nonumber\\
a_{1,\,0,\,0}^{(1)} = -\sigma^4\left(\sigma^8-109\sigma^6+517\sigma^4+217\sigma^2+270\right),\nonumber\\
a_{0,\,0,\,1}^{(1)} = \sigma^4\left(\sigma^8+20\sigma^6-158\sigma^4-28\sigma^2-27\right),\nonumber
\end{gather}
\begin{gather*}
a_{1,\,0,\,0}^{(2)} = -2\sigma^3\bigl(3\sigma^{10}+18\sigma^8-146\sigma^6\\-172\sigma^4+183\sigma^2-270\bigr),\\
a_{0,\,0,\,1}^{(2)} = 2\sigma^3\bigl(3\sigma^{10}-63\sigma^8+314\sigma^6\\-218\sigma^4+19\sigma^2+9\bigr),\\
a_{1,\,0,\,0}^{(3)} = 2\sigma^2\left(\sigma^2-1\right)\bigl(7\sigma^{10}-58\sigma^8+38\sigma^6\\+52\sigma^4-181\sigma^2+270\bigr),\\
a_{0,\,0,\,1}^{(3)} = -2\sigma^2\left(\sigma^2-1\right)\bigl(7\sigma^{10}-79\sigma^8+282\sigma^6\\-154\sigma^4-\sigma^2+9\bigr),\\
a_{1,\,0,\,0}^{(4)} = \sigma\left(1-\sigma^2\right)^3\left(11\sigma^8-99\sigma^6-61\sigma^4+7\sigma^2+270\right),\\
a_{0,\,0,\,1}^{(4)} = \sigma\left(\sigma^2-1\right)^3\left(11\sigma^8-48\sigma^6+66\sigma^4+8\sigma^2+27\right),\\
a_{1,\,0,\,0}^{(5)} = \left(\sigma^2-1\right)^5\left(3\sigma^6-20\sigma^4-21\sigma^2+54\right),\\
a_{0,\,0,\,1}^{(5)} = -\left(\sigma^2-1\right)^5\left(3\sigma^6+7\sigma^4-11\sigma^2+9\right).
\end{gather*}
The coefficients $a_3$, $a_{0,\,0,\,0}$, $a_{1,\,0,\,0}$, and
$a_{0,\,0,\,1}$ are all real and depend only on one dimensionless
parameter $kh$. Their behavior as functions of $kh$ is shown in Fig.
\ref{fig:an}. It can be seen that Eq.~(\ref{eq:HONLSE}) is valid at
$kh \gtrsim 1$, where its coefficients do not diverge.  At
$kh\rightarrow\infty$, the following asymptotics are easily
obtained:
\begin{equation*}
a_3 = \frac{1}{4},\;\; a_{0,\,0,\,0} = -\frac{1}{2},\;\;
a_{1,\,0,\,0} = \frac{3}{2},\;\; a_{0,\,0,\,1} = \frac{1}{4}.
\end{equation*}
They coincide with the corresponding coefficients of Dysthe's
equation \cite{Trulsen_Dysthe_2000}, except for the term including
the wave-induced mean flow. This term cannot be explicitly
reconstructed from Eq.~(\ref{eq:HONLSE}) because of the additional
power expansion of the wave-induced mean flow made to derive this
equation. However, it can be reconstructed from the equations
generating Eq.~(\ref{eq:HONLSE}), at the stage when the wave-induced
mean flow has not been excluded from the equation for $u$
\cite{SedletskyJETP2003}. By setting
$a_3=a_{1,\,0,\,0}=a_{0,\,0,\,1}\equiv0$, Eq.~(\ref{eq:HONLSE}) is
reduced to the standard NLSE derived by Hasimoto and
Ono~\cite{Hasimoto_Ono_1972}.


The dimensionless free surface displacement is expressed in terms of
$u$ as follows
\begin{equation}\label{Eq:zeta}
k\eta =\alpha_0|u|^2 + \alpha_1\Re\bigl( u\exp(\I\theta)\bigr) + 2\alpha_2\Re\bigl( u^2\exp(2\I\theta)\bigr),
\end{equation}
\begin{equation*}
\alpha_0 =\frac{2\left(1-\sigma^2\right)kh+\sigma}{c\nu},\quad\alpha_1 = \frac{1}{\sqrt{c}},\quad\alpha_2=\frac{3-\sigma^2}{8c\sigma^3},
\end{equation*}
where $\theta = kx - \omega t = \chi - c\tau$ is the wave phase. The
similar expression for the velocity potential is given in
Ref.~\cite{UJP_2014}.

The region of modulation instability of a homogeneous solution
$u(\chi,\,\tau)=u_0\exp\left(\I a_{0,\,0,\,0}|u_0|^2\tau\right)$ to
Eq.~(\ref{eq:HONLSE}) can be determined by introducing a small
perturbation to the amplitude $u_0$ (as described, e.g., in Refs.
\cite{Dysthe_1979,Ablowitz_2001}):
\begin{gather}
u(\chi,\,\tau)=\bigl(u_0+\epsilon(\chi,\,\tau)\bigr)\exp\left(\I a_{0,\,0,\,0}|u_0|^2\tau\right),\\
\epsilon(\chi,\,\tau)=\epsilon_0^{+}\exp\left(\I\kappa\chi-\I\Omega\tau\right)+\epsilon_0^{-}\exp\left(-\I\kappa\chi+\I\Omega^*\tau\right).\nonumber
\end{gather}
Here, we assume the perturbation frequency $\Omega$ to be
complex-valued and the perturbation wave number $\kappa$ to be real.
Substituting this ansatz in Eq.~(\ref{eq:HONLSE}) leads to the
following dispersion relation between $\Omega$  and $\kappa$:
\begin{multline}\label{eq:MI_Omega}
\Omega = \left(a_1+ a_{1,\,0,\,0}|u_0|^2\right)\kappa + a_3 \kappa^3 \\
\pm \kappa\sqrt{2a_2 a_{0,\,0,\,0}|u_0|^2 + a_{0,\,0,\,1}^2|u_0|^4 + a_2^2\kappa^2}.
\end{multline}
A homogeneous solution is modulationally unstable when the
perturbation exponentially grows with time. This happens when
$\Im\Omega>0$, which effectively requires the radicand in Eq.
(\ref{eq:MI_Omega}) to be negative. This condition is satisfied when
\begin{equation}\label{eq:MI}
a_{0,\,0,\,0} < - a_{0,\,0,\,1}^2 |u_0|^2,
\end{equation}
where we took into account that $a_2 = \frac{1}{2}$. In the NLSE
case, when $a_{0,\,0,\,1}\equiv 0$, condition (\ref{eq:MI}) is
reduced to the well-known modulation instability criterion
$a_{0,\,0,\,0} < 0$, which holds true at $kh>1.363$. In the HONLSE
case, this threshold slightly shifts to higher $kh$, depending on
the value of $u_0$.

Bright and dark solitons are defined as \cite{Kivshar_1998}
\begin{align}
u_B(\chi,\,\tau)\!=&\,u_0\sech\bigl(K(\chi-V\tau)\bigr)\exp\left(\I\kappa\chi-\I\Omega\tau\right),\label{eq:soliton_bright}\\
u_D(\chi,\,\tau)\!=&\,u_0\tanh\bigl(K(\chi-V\tau)\bigr)\exp\left(\I\kappa\chi-\I\Omega\tau\right).\label{eq:soliton_dark}
\end{align}
Here, $u_0$, $\kappa$, $\Omega$, and $V$, are the soliton's complex
amplitude, wave number, frequency, and speed, respectively. The
soliton's initial position was chosen to be located at $\chi = 0$.
In the NLSE case, bright one-soliton solutions of form
(\ref{eq:soliton_bright}) were first derived in Ref.
\cite{Zakharov_1971}. The following relationships between the
soliton parameters can be established in this case:
\begin{gather}
K = |u_0|\sqrt{-a_{0,\,0,\,0}},\nonumber\\
\Omega = \kappa a_1 + \frac{1}{2}\left(K^2-\kappa^2\right),\; V = a_1 -\kappa.\label{eq:BS_NLSE}
\end{gather}
Bright NLSE solitons can exist at $a_{0,\,0,\,0} < 0$, which exactly
corresponds to the region where a homogeneous solution is
modulationally unstable. The parameters $u_0$ and $\kappa$ are free
parameters of the problem. However, of physical relevance are only
such amplitudes and wave numbers that preserve the conditions of
small amplitude and narrow spectrum: $|u_0| \ll 1$, $\kappa \ll 1$.

Dark solitons of form (\ref{eq:soliton_dark}) are referred to as
fundamental dark solitons \cite{Kivshar_1998}. In the NLSE context,
they were first derived in a more general form in Ref.
\cite{Zakharov_1973}. Dark NLSE solitons exist below the modulation
instability threshold, at $a_{0,\,0,\,0} > 0$, and the corresponding
relationships between the parameters are
\begin{gather}
K = |u_0|\sqrt{a_{0,\,0,\,0}},\nonumber\\
\Omega = \kappa a_1 -\frac{1}{2}\left(2K^2+\kappa^2\right),\; V = a_1 -\kappa.\label{eq:DS_NLSE}
\end{gather}

\begin{figure}[t]
\includegraphics[width=\columnwidth]{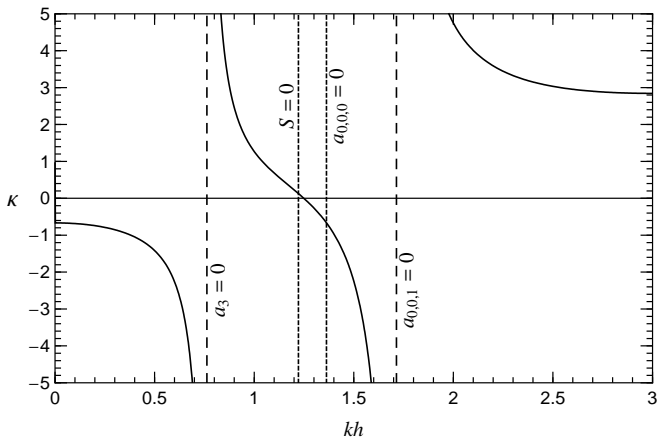}
\caption{\label{fig:kappa_kh} Wave number $\kappa$ of the HONLSE
soliton versus $kh$.}
\end{figure}

\begin{figure}[t]
\includegraphics[width=\columnwidth]{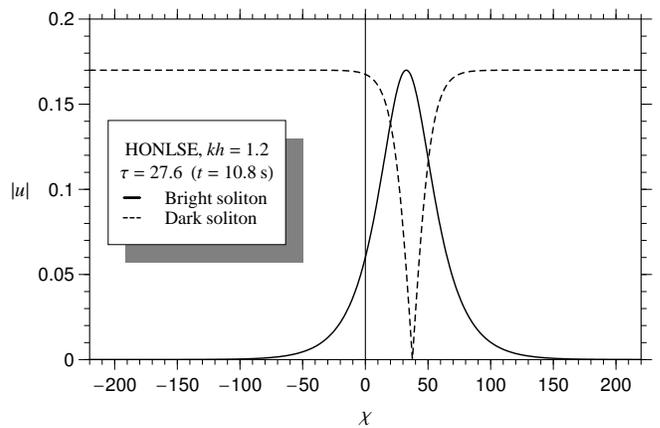}
\caption{\label{fig:BDsoliton} Bright and dark solitons of HONLSE (\ref{eq:HONLSE}) for $kh=1.2$ and $u_0 = 0.17$.
The bright soliton parameters are determined by relations (\ref{eq:BS_HONLSE}): $K \approx 0.052$, $\kappa \approx 0.23$, $\Omega \approx 0.30$, $V\approx 1.18$.
The dark soliton parameters are chosen such that they closely correspond to the water-tank experiments described in Ref.~\cite{Chabchoub_PRL_2013_Dark}:
$K \approx 0.066$, $\kappa = 0$, $\Omega \approx -0.0043$, $V=a_1\approx 1.39$.}
\end{figure}

\begin{figure}[t]
\includegraphics[width=\columnwidth]{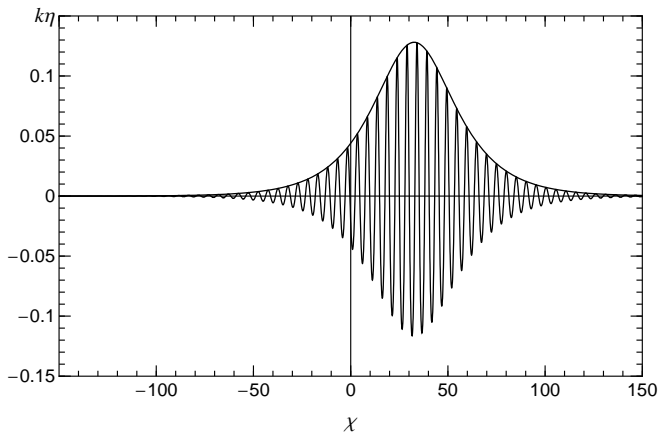}
\caption{\label{fig:BS_wave} Free surface profile with envelope corresponding to the bright soliton shown in Fig. \ref{fig:BDsoliton} ($kh=1.2$).}
\end{figure}

Relations (\ref{eq:BS_NLSE}) and (\ref{eq:DS_NLSE}) are no longer
valid in the HONLSE case, when the high-order terms are not set
equal to zero. In this case, bright NLSE solitons are transformed
into quasi-solitons, which were obtained by numerical integration of
Eq.~(\ref{eq:HONLSE}) in Ref.~\cite{UJP_2014}. Below we prove that
HONLSE~(\ref{eq:HONLSE}) admits a new family of exact bright soliton
solutions that exist below the modulation instability threshold
$kh\approx 1.363$. Substituting ansatz (\ref{eq:soliton_bright}) in
Eq.~(\ref{eq:HONLSE}), one can obtain the following relations:
\begin{subequations}\label{eq:BS_HONLSE}
\begin{gather}
K = |u_0|\sqrt{S},\quad S = -\frac{a_{1,\,0,\,0}+a_{0,\,0,\,1}}{6a_{3}},\label{eq:K_HONLSE}\\
\kappa = \frac{a_{1,\,0,\,0}+a_{0,\,0,\,1}-6a_{3}a_{0,\,0,\,0}}{12a_{3}a_{0,\,0,\,1}},\label{eq:kappa_HONLSE}\\
\Omega = \kappa a_1+\frac{1}{2}\Bigl(K^2\left(1-6\kappa a_{3}\right)-\kappa^2\left(1-2\kappa a_{3}\right)\Bigr),\\
V = a_1 -\kappa + \left(3\kappa^2 - K^2\right) a_{3}.
\end{gather}
\end{subequations}
Note that such relations were derived earlier for the general form
of Eq.~(\ref{eq:HONLSE}) in Refs.
\cite{Potasek_Tabor_1991,Gromov_1999,Karpman_2001} without referring
to water waves. Anzats (\ref{eq:soliton_bright}) is an exact
solution to Eq.~(\ref{eq:HONLSE}) when the radicand in
Eq.~(\ref{eq:K_HONLSE}) is positive, $S>0$. This condition holds
true in the following depth range:
\begin{equation}
0.763\lesssim kh\lesssim1.222,
\end{equation}
which lies below the modulation instability threshold. The soliton
wave number $\kappa$ is not a free parameter any longer. Its
behavior as a function of $kh$ is shown in Fig.~\ref{fig:kappa_kh}.
The condition of narrow spectrum (slow modulation) holds true only
in a narrow range around $kh\approx 1.249$, where $\kappa=0$. At
$kh=1.2$, we have $\kappa\approx 0.23$. This wave number can still
be regarded small as compared to unity, but not so small as in the
NLSE case. Figure~\ref{fig:BDsoliton} shows a bright soliton
solution to HONLSE (\ref{eq:HONLSE}) along with the dark soliton
that exists at the same dimensionless depth $kh=1.2$. The
corresponding free surface profile computed by formula
(\ref{Eq:zeta}) is shown in Fig.~\ref{fig:BS_wave}. The parameters
of the dark soliton shown in Fig.~\ref{fig:BDsoliton} were selected
such that they closely correspond to the water-tank geometry and
dark-soliton experiments described in Ref.
\cite{Chabchoub_PRL_2013_Dark}. These experiments were performed at
the same depth ($kh = 1.2$), the carrier wave number was
$k=3$~m$^{-1}$, and the carrier amplitude was $a = 0.04$ m. The
corresponding dimensionless amplitude $ka = 0.12$ corresponds to the
amplitude of soliton shown in Fig. \ref{fig:BS_wave}. The evolution
time $\tau=27.6$ ($t = 10.8$ s) was chosen such that the wave group
travels the distance $L = V_g\, t = 12.8$~m ($V_g = 1.19$~m/s) to
the last wave gauge position along the wave tank in the experimental
setup \cite{Chabchoub_PRL_2013_Dark}.

The dark soliton shown in Fig.~\ref{fig:BDsoliton} was computed by
integrating HONLSE (\ref{eq:HONLSE}) numerically using the
split-step Fourier technique described in \cite{UJP_2014} with the
dark NLSE soliton taken as the initial wave form. On the adopted
time scales, the corresponding numerical HONLSE solution identically
followed the initial dark soliton shape. Deviations from NLSE can be
observed on much larger time scales ($\tau\gtrsim2000$) in the form
of slight oscillations of the wave envelope. Besides such dark
quasi-solitons, HONLSE (\ref{eq:HONLSE}) admits a family of exact
dark soliton solutions defined by the following set of
parameters:\pagebreak[2]
\begin{subequations}
\begin{gather}
K = |u_0|\sqrt{-S},\\
\Omega = \kappa a_1 + K^2\left(6\kappa a_{3}-1\right)-\frac{\kappa^2}{2}\left(1-2\kappa a_{3}\right),\\
V = a_1 -\kappa + \left(3\kappa^2 + 2K^2\right) a_{3},
\end{gather}
\end{subequations}
with $\kappa$ given by the same formula (\ref{eq:kappa_HONLSE}) as
for the bright solitons. These relations can be obtained in the same
way as relations (\ref{eq:BS_HONLSE}). Similar expressions for dark
HONLSE solitons were also derived in Ref.~\cite{Potasek_Tabor_1991}
without referring to water waves. Dark HONLSE solitons exist at
$S<0$, i.e., at $kh\gtrsim 1.222$. In contrast to bright solitons,
NLSE and HONLSE dark solitons can exist in the overlapping depth
ranges.

Thus, we presented for the first time new families of bright and
dark solitons that exist below the modulation instability threshold
for water waves on the surface of ideal finite-depth fluid. These
solitons are the exact solutions to the HONLSE derived earlier in
Ref.~\cite{UJP_2014} to describe the evolution of slowly modulated
wave trains. The results of this study can readily be used in
water-tank experiments to supplement the recent observations of dark
solitons described in Ref.~\cite{Chabchoub_PRL_2013_Dark}.

We would like to thank Dr. Denys Dutykh for helpful discussions.

\end{document}